# Arp's Indomitable Universe

Domingos S. L. Soares
Physics Department
Federal University of Minas Gerais – UFMG
C. P. 702, 30123-970 Belo Horizonte, MG, Brazil
dsoares@fisica.ufmg.br –
www.fisica.ufmg.br/~dsoares

Marcos C. D. Neves
Physics Department
State University of Maringá – UEM
87020-900 Maringá, PR, Brazil
macedane@yahoo.com – www.galileo-400-anos.blogspot.com

Andre K. T. Assis
Institute of Physics 'Gleb Wataghin'
University of Campinas — UNICAMP
13083-859 Campinas, SP, Brazil
assis@ifi.unicamp.br – www.ifi.unicamp.br/~assis

We present some aspects of the work and personality of Halton Christian Arp (1927-2013).

We are used to being astonished by beautiful images of galaxies, splendorous arrangements of stars, gas and interstellar dust. The Sun and its planetary system are located in one of them, the Milky Way galaxy, a giant spiral galaxy, with the Sun sitting in the periphery of its disk. Our similar-sized neighbor is the Andromeda galaxy, another spiral, formidable and harmonious, especially when contemplated with a telescope. Incidentally, the Andromeda galaxy is the most distant cosmic object visible with the unaided eye.

But galaxies, like people, can also be complicated and strange, and interact in a very peculiar way. And that is what we shall deal with here – with such a complex and, to a degree, indomitable universe.

One of the first astronomers to embrace the "weird" galaxies was the American Halton Christian Arp[*] (1927-2013), born in New York City, United States, and well-known among his friends and close colleagues by the nickname "Chip." He graduated in astrophysics in 1953, but even before this date, he had worked under Edwin Hubble (1889-1953). In 1957 he began working at the Palomar Observatory, which then had the largest telescope in the world, the 200 inch Hale telescope. Arp worked for 29 years at Palomar until he finally moved to the Max Planck Institute for Astrophysics, in Germany in 1983, where he stayed active in astronomy until his death in 2013.

Our story begins with Arp's interest in galaxies with strange shapes, in the majority of the cases caused by gravitational interactions with their close neighbors. Some of them are clearly in the process of merging. Two or more galaxies interact so strongly that their original shapes are destroyed and the interacting group quickly turns into a single object made up of the aggregation of stars and all material present in the original galaxies. As soon as he began his work at Palomar, he started to look for galaxies of this kind. After quite some time, in 1966 he published the result: an atlas of "weird" galaxies, the *Atlas of Peculiar Galaxies*, which, to this day, is still used in astronomical research dedicated to understanding the evolution of galaxies.

The Atlas consists of a set of 338 sky fields, presented in rectangular format, identified by "Arp nnn," where "nnn" is the number of the field. Each field shows one or more galaxies that always exhibit some peculiar structure, often caused by the gravitational interaction between them. The field sizes range from – the smallest – 2 arcminutes wide, up to 1 degree wide (Arp 318). The fields were observed with the 200 inch telescope and with the 48 inch Schmidt telescope, the latter well suited for the larger sky fields. The images were engraved on photographic emulsions specially designed for astronomical use.

---

[*] After being ill for some time, Halton *Chip* Arp passed away on December 28, 2013 in Munich, Germany. He was born on March 21, 1927 in New York City, United States. Arp is well known for his non orthodox views in astrophysics and especially in cosmology. He is the author of the renowned *Atlas of Peculiar Galaxies* (1966), of innumerable scientific publications and of the books *Quasars, Redshifts and Controversies* (1987) and *Seeing Red: Redshifts, Cosmology and Academy* (1998). Arp was living in Germany since 1983 where he was a researcher at the Max Planck Institute for Astrophysics. According to a communication from our colleague Hilton Ratcliffe "His legacy will live on with anyone who is concerned with objective scientific reason. Chip was a fine man and a very dear friend and colleague. His final paper is currently with the referees of *ApJ*, and I shall let you all know when it is published."

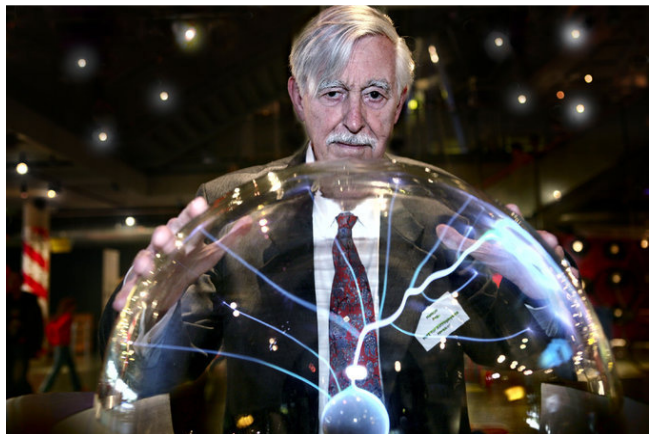

Figure 1 – Halton "Chip" Arp. [Overbye, 2014]

Arp sorted the general features of his *Atlas* into four large groups.
- Arp 1 to 101 – Spiral galaxies that have: low brightness, subdivided spiral arms, arms with detached segments, three arms, one arm, faint companions and elliptical galaxies as companions.
- Arp 102 to 145 – Elliptical galaxies with the following features: connected to spirals, close to disturbed spirals, close to galaxy fragments and close to apparently ejected material.
- Arp 146 to 268 – Galaxies (not included in previous groups) that have rings, jets, filaments, tails, amorphous spiral arms and loops adjacent to the main body.
- Arp 269 to 338 – Double galaxies connected by stellar arms (bridges) and with long filaments, galaxy groups and galaxy chains.

Photographic illustrations of these groups, obtained by Arp, are available online at:
http://nedwww.ipac.caltech.edu/level5/Arp/frames.html.

The field Arp 81, shown in Figure 2, is a pair of spiral galaxies which began colliding, *i.e.*, at closest approach, 100 million years ago. During this period, the original shapes of the galaxies became very disturbed. Their original nuclei and signs of what were their stellar disks are readily visible. At the top of the image is a "tail" produced by the gravitational force of one galaxy over the stellar disk of the other, in this case, the largest one NGC 6621, to the left. Such a structure is called a "tidal tail." The formation of this tail is due to the very same phenomenon that produces ocean tides on Earth, from which it gets its name. Ocean tides are caused by the com-bined gravitational forces of the Moon – mainly this – and of the Sun on the terrestrial oceans.

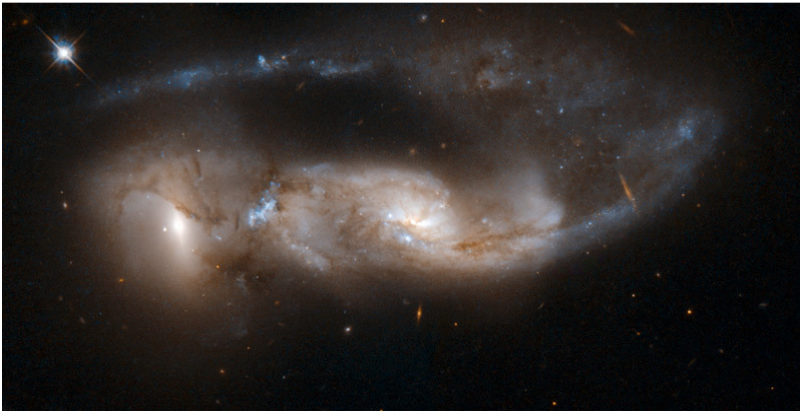

Figure 2 – The field Arp 81, a pair of galaxies in strong gravitational interaction. The galaxy on the left is NGC 6621 and on the right is NGC 6622. Arp 81 is 300 million light years from us. One of the main consequences of these kind of interactions is an increase in star formation in the galaxies (Credit: Hubble Space Telescope/NASA).

The field Arp 244, shown in Figure 3, is another example where tidal tails clearly show up. This system is known as the *Antennae Galaxies* because of its peculiar shape. The tails are definitely formed by the gravitational interaction between stars belonging to the disks of the two spiral galaxies that form the system.

This was demonstrated for the first time by the brothers Alar Toomre, astronomer and mathematician, and Juri Toomre, astronomer, by means of numerical simulations performed with a computer in the late 1960s. Alar and Juri Toomre were born in Estonia. In 1949 they immigrated to the United States where they graduated and furthered their astronomical research.

Arp 244's galaxies were represented by stellar disks, each one with 120 particles. These particles, which represent the galaxy's stars, undergo the action of the gravitational force exerted by both galaxies. Initially the galaxies were perfect disks, and then they were "thrown" into in a mutual elliptical orbit. Through the action of the gravitational forces, stars gradually abandon the original disk-like distribution in the subsequent orbital motion. As illustrated in Figure 4, some of the stars move far away from the majority of stars that formed the disks. The Toomre brothers chose an appropriate angle of view of the simulated system in order to obtain an image that is similar to the real system.

The simulations, of this system and others, performed by the Toomres showed, for the first time in the history of astronomy, that the shapes of interacting galaxies were the result of gravitational interactions among the constituents of the interacting system. Astronomy, the study of the locations of celestial bodies in space and time, is here supplemented by a physical law – Newton's law of universal gravitation – and becomes authentic astrophysics.

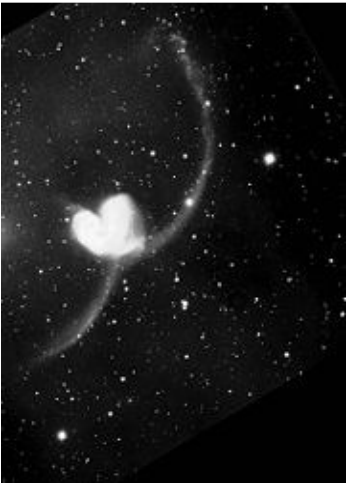

Figure 3 – The field Arp 244, another pair in strong interaction (NGC 4038 and NGC 4039). The tidal tails that emerge from both galaxies in the pair give the impression of insect antennae, hence the popular name of this system, "Antennae Galaxies" (Credit: Brad Whitmore/STScI).

After the publication of Alar and Juri Toomre's investigation in *ApJ*, in 1972, the study of interacting galaxies by means of numeric-computational simulations became a topic of great interest in the astronomical community. This kind of simulation is known as "N body simulation," where the systems involved in the interactions are represented by systems with N particles. In the simulation of Arp 244, N = 240. Nowadays, with the development of supercomputers, N can easily reach 100 thousand. Furthermore, galaxies are represented by more realistic models with the inclusion of gas and dust, which are important components of many galaxies. Other processes are also considered such as the chemical evolution of galaxies and the consequent changing in their stellar populations, and hydrodynamic interactions, typical of gaseous components.

The indomitable universe cannot only be seen in the cosmic dimension. It also marked Arp's attitude throughout his entire scientific life.

He is known for his strong opposition to the standard model of cosmology. The orthodox scientific community acclaims the standard model, the Big-Bang model. Nevertheless, its critics point to its many inconsistencies, especially when the predictions of the theory are confronted with astronomical observations. Most of Arp's ideas in this and other scientific topics were publicized, for the specialist and non-specialist, in his three books, *Quasars, Redshifts and Controversies* (1987), *Seeing Red* (1998), and *Catalogue of Discordant Redshift Associations* (2003).

In 2003 one of the authors, MCDN, met Halton Arp twice in Italy. The first meeting happened in Naples, at the Istituto Italiano per gli Studi

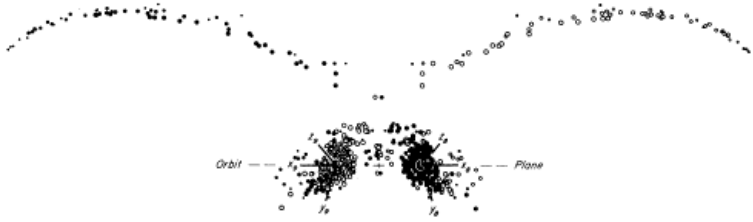

Figure 4 – Result of the numerical simulation of the Antennae Galaxies. Filled little circles represent stars of one galaxy and open circles stars of the other. The orbital plane of the original galaxies, initially disks of stars, is indicated by "Orbital Plane." The mutual gravitational forces act in such a way that they pull disk stars out to form tidal tails. Compare this with the previous figure, which is an image of the real system (Credit: Alar Toomre and Juri Toomre).

Filosofici, at Palazzo Serra di Cassano, during the International Congress on Science and Democracy, organized by Marco Mamone Capria.* The second meeting took place in Pavia at the International Workshop on Alternative Cosmologies, organized by Enrico Giannetto.

It was a great pleasure for MCDN to meet him. Three years later he invited Arp to write a paper for the *Acta Scientiarum* journal of the Universidade Estadual de Maringá (Brazil), as MCDN was responsible for the section on natural sciences of this publication. He accepted the invitation and published the paper "Cosmology and Physics."† In this work he considered the observation of extra-galactic bodies, showing that their redshifts were a function of their age, instead of being a function of their recessional velocities. Arp believed that a model like the big bang was not necessary in order to understand the behavior of the universe. In this paper he pointed out that the analysis of new data indicated the need for a new physics in order to satisfy experiments which had up to now been based on Einstein's theory of relativity (TR). He believed that the present situation requires simple logic rather than mathematical formulae for new solutions. It also requires new rules which are outside the normal scope of TR. He criticized the paradigms which confine science to the news media by an inadequate use of an old physics which is not open to the new possibilities required for a deep understanding of nature and the universe.

He concluded this paper as follows:

> Probably many of these independent researchers have wondered whether Academia is a doomed institution. Has 800 years of uncritical approval and acceptance led to senescence? Has a self-serving feedback loop between academia and the news media convinced the two parties that fundamental assumptions can never be questioned?

---

\* [Fanuzzi, Gargano and Chiaro, 2010].

† [Arp, 2000].

What is to be done? Press with logic and insistence on both estates? Of course! Press ahead with independent research and mutual support from all such researchers. Of course! It is my hope that by putting forth the candid thoughts of this paper, we may use each other's results and concepts to unify science on all scales, and across all disciplines, in a way that will lead to more fruitful discussions and understandings in the future.

Below we present a few of the events which took place in Italy, organized by the Istituto Italiano per gli Studi Filosofi,[*] in which Arp participated. The participants are also listed to give an idea of the persons with whom he was probably in contact during these meetings.

- **NEW IDEAS IN ASTRONOMY,** In collaborazione con l'Istituto Veneto di Scienze, Lettere ed Arti, il Dipartimento di Astronomia dell'Università di Padova e l'Osservatorio Astronomico di Padova. Venezia, 5-7 maggio 1987. Relazioni di: C. Maccagni, F. Hoyle, J. Heidmann, R. Kraft, A. Renzini, C. Chiosi, L. Rosino, V. Ambartsumian, R. Dickens, M. Roberts, R. Wolstencroft, S. Bonometto, M. Capaccioli, G. Bertin, M. Burbidge, S. Di Serego-Alighieri, E. Khachikian, P. Rafanelli, I. Pronik, J. Sulentic, N. Sharp, G. Schnur, S. Cristiani, H. Arp, W. Tifft, W. Napier, W. Saslaw, W. Alfven, L. Woltjer, G. Burbidge, A. Treves, A. Zensus, J. Narlikar, J. P. Vigier, G. Börner, R. Sanders, J. C. Pecker, K. Rudnicki, J. Wampler, W. Brinkmann, A. Cavaliere, D. Sciama, V. Clube, R. Ruffini.
- **KOSMOS: LA COSMOLOGIA OGGI TRA FILOSOFIA E SCIENZA (COSMOS: COSMOLOGY BETWEEN PHILOSOPHY AND SCIENCE),** In collaborazione con l'Istituto Gramsci Veneto e il Goethe Institut), Venezia, 8-9 maggio 1987. Relazioni di Umberto Curi, Livio Gratton, Halton C. Arp, Dennis W. Sciama, Jayant V. Narlikar, Enrico Bellone, Geoffrey Burbidge, Jean-Pierre Vigier, Oddone Longo, Nicola Badaloni, Dieter Wandschneider, Fred Hoyle, Carlo Sini, Jean Heidmann, Paolo Zellini, Jean-Claude Pecker.
- **IL PRINCIPIO ANTROPICO (THE ANTHROPIC PRINCIPLE),** In collaborazione con l'Istituto Gramsci Veneto, il Goethe Institut, il Dipartimento di Astronomia dell'Università di Padova), Venezia, 18-19 novembre 1988. Relazioni di: John Barrow, Oddone Longo, Brandon Carter, Hubert Reeves, Fred Hoyle, Livio Gratton, Dennis W. Sciama,

---

[*] [Fanuzzi, Gargano and Chiaro, 2010] and [Gargano, 2005].

Jean Heidmann, Friedrich Cramer, Nicola Dalla Porta, Halton C. Arp, George Coyne, Bernulf Kanitscheider, Massimo Cacciari.

- **LE ORIGINI DELL'UNIVERSO (ORIGIN OF THE UNIVERSE)**, In collaborazione con: Istituto Gramsci Veneto, Dipartimento di Astronomia dell'Università di Padova, Goethe Institut. Venezia, 15-16 dicembre 1989. Relazioni di: Umberto Curi, Rudolf Kippenhahn, Ferruccio Franco Repellini, George Ellis, Roberto Barbon, Giulio Giorello, Marco Senaldi, Livio Gratton, Volker Weidemann, Remo Ruffini, Remo Bodei, Paolo Rossi, Paul Davies, Roger Penrose, Martin Rees, Halton C. Arp, Juan Casanovas S. J., Jean Heidmann, Dennis W. Sciama, Jean-Pierre Vigier.

- **ORIGINI: L'UNIVERSO, LA VITA, L'INTELLIGENZA (ORIGINS: UNIVERSE, LIFE, INTELIGENCE)**, In collaborazione con l'Istituto Gramsci Veneto e il Dipartimento di Astronomia dell'Università di Padova. Venezia, 18-19 dicembre 1992. Relazioni di: Umberto Curi, Oddone Longo, Paolo Budinich, Margherita Hack, Massimo Calvani, Julian Chela-Flores, Cristiano Cosmovici, André Brack, Francesco Bertola, Jean Heidmann, Reginaldo Francisco O. P., Dennis W. Sciama, Halton Arp.

- **LA BELLEZZA DELL'UNIVERSO (THE BEAUTY OF THE UNIVERSE)**, In collaborazione con l'Istituto Gramsci Veneto e l'Università di Padova. Venezia, 17-18 dicembre 1993. Relazioni di: Umberto Curi, Carlo Sini, Nicolò Dallaporta, Giangiorgio Pasqualotto, Giò Pomodoro, Massimo Calvani, Peter Kafka, Giovanni Boniolo, Bruno Bertotti, Halton Arp, Francesco Bertola, Werner Zeilinger, Enrico Bellone, Paolo Bettiolo, Jean Heidmann, Franco Rella, Dennis Sciama.

- **CONVEGNO INTERNAZIONALE: SCIENZA E DEMOCRAZIA (INTERNATIONAL CONGRESS: SCIENCE AND DEMOCRACY)**, In collaborazione con l'Università di Perugia. 12-14 giugno 2003. Relazioni di: Stefano Dumontet, Marco Mamone Capria, David Rasnick, Marcos Cesar Danhoni Neves, T. Tonietti, A. Drago, G. Moran, Halton Arp, David Rasnick, Roberto Germano, Anthony Liversidge, Sergio Siminovich, Frank Lad, Marinella Leo, Raffaele Capone, Marco Mamone Capria, Sergio Calderaro, Adriana Valente, I. Nobile, Pasquale Santé, Federico Di Trocchio.

- **PAVIA INTERNATIONAL WORKSHOP ON ALTERNATIVE COSMOLOGIES**, Organized by Eric Lerner and Enrico Giannetto, June 23-25, 2003, Università degli Studi di Pavia. Relazioni di: Jack Sulentic, Sisir Roy, Menas Kafatos, Chuck Gallo, Anthony Peratt, Hal-ton Arp, Ya. Baryshev, H. C. Kandpal, Eric J. Lerner, Jacques Moret-Bailly, Marcos Cesar Danhoni Neves, Georges Paturel, Francesco Sylos Labini.

In honor of his brilliant work and the bravery he always showed in his constant fight against the establishment, two of the authors (AKTA and DSLS) have done a Portuguese translation of one of his books.[*]

Randall Meyers, the celebrated composer, especially known for the music he composed for the Hollywood film "The English Patient," who directed and produced the documentary film "Universe: The Cosmology Quest,"[†] published on his official website[‡] a precious short film in memory of Fred Hoyle, showing his meeting with Chip Arp. In this encounter we perceive the power of anti-paradigmatic arguments against the *establishment.* It is vitally important to keep alive the memory of this great warrior of modern science.

---

[*] [Arp, 1998] and [Arp, 2001].

[†] [Meyers, 2004].

[‡] [Meyers, 2012].